\documentclass[aps,prb,a4paper,twocolumn,noshowpacs,superscriptaddress,floatfix]{revtex4-1}
\usepackage{amsmath, amssymb}
\usepackage{graphicx}
\usepackage{times}
\usepackage{dcolumn}
\usepackage{array}
\usepackage{threeparttable}
\usepackage{setspace}
\usepackage{multirow}
\usepackage{booktabs}
\emergencystretch=\maxdimen
\newcommand{\LOFBS}{$\mathrm{La}\mathrm{O_{0.5}}\mathrm{F_{0.5}}\mathrm{BiS_{2}}$}

\newcommand{\COFBS}{$\mathrm{Ce}\mathrm{O_{0.5}}\mathrm{F_{0.5}}\mathrm{BiS_{2}}$}
\newcommand{\NOFBS}{$\mathrm{Nd}\mathrm{O_{0.5}}\mathrm{F_{0.5}}\mathrm{BiS_{2}}$}

\newcommand{\LnOFBS}{$Ln\mathrm{O_{0.5}}\mathrm{F_{0.5}}\mathrm{BiS_{2}}$}
\newcommand{\msr}{$\mu$SR}
\newcommand{\BiS}{$\mathrm{BiS_{2}}$}
\newcommand{\BOS}{$\mathrm{Bi_{4}O_{4}S_3}$}

\begin{document}

\title{Superconducting gap structure in ambient-pressure-grown \bf{\LOFBS~}}

\author{Jian Zhang}
\author{K. Huang}
\author{Z. F. Ding}
\affiliation{State Key Laboratory of Surface Physics, Department of Physics, Fudan University, Shanghai 200433, People's Republic of China}
\author{D. E. MacLaughlin}
\affiliation{Department of Physics and Astronomy, University of California, Riverside, California 92521, USA}
\author{O. O. Bernal}
\affiliation{Department of Physics and Astronomy, California State University, Los Angeles, California 90032, USA}
\author{P. -C. Ho}
\affiliation{Department of Physics, California State University, Fresno, California 93740,USA}
\author{C. Tan}
\affiliation{State Key Laboratory of Surface Physics, Department of Physics, Fudan University, Shanghai 200433, People's Republic of China}
\author{X. Liu}
\affiliation{State Key Laboratory of Surface Physics, Department of Physics, Fudan University, Shanghai 200433, People's Republic of China}
\author{D. Yazici}
\affiliation{Department of Physics and Center for Advanced Nanoscience, University of California, San Diego, La Jolla, California 92093,USA}
\author{M. B. Maple}
\affiliation{Department of Physics and Center for Advanced Nanoscience, University of California, San Diego, La Jolla, California 92093,USA}
\author{Lei Shu}
\altaffiliation[Corresponding Author: ]{leishu@fudan.edu.cn}
\affiliation{State Key Laboratory of Surface Physics, Department of Physics, Fudan University, Shanghai 200433, People's Republic of China}
\affiliation{Collaborative Innovation Center of Advanced Microstructures, Nanjing 210093, People's Republic of China}

\date{\today}

\begin{abstract}
We have performed transverse-field muon spin relaxation (TF-\msr) measurements on ambient-pressure-grown polycrystalline \LOFBS. From these measurements, no signature of magnetic order is found down to 25 mK. The value of the magnetic penetration depth extrapolated to 0 K is 0.89 (5) $\mu$m. The temperature dependence of superconducting penetration depth is best described by either a multigap $s + s$~-wave model with $\Delta_{1}$ = 0.947 (7) meV and $\Delta_{2}$~=~0.22~(4)~meV or the ansiotropic $s$-wave model with $\Delta(0)$~=~0.776~meV and anisotropic gap amplitude ratio $\Delta_{min}/\Delta_{max}$~=~0.34. Comparisons with other potentially multigap \BiS-based superconductors are discussed. We find that these \BiS-based superconductors, including $\mathrm{Bi_{4}O_{4}S_3}$ and the high-pressure synthesized \LOFBS, generally conform to the Uemura relation.

\end{abstract}
\maketitle

\section{Introduction}
The discovery of superconductivity in the \BiS~layered compounds \LnOFBS~($\mathit{Ln}$ = La, Ce, Pr, Nd, Yb) and $\mathrm{Bi_{4}O_{4}S_3}$, with the highest $T_{c}$ = 10.6 K observed in the La-member, has attracted considerable attention \cite{YoshikazuPRB.86.220510,Yazici13,Yazici2015218,YoshikazuJPSJ.81.114725}. In this new family, the superconductivity arises from the \BiS~layers, analogous to the $\mathrm{CuO_{2}}$ layers in the high-$T_{c}$ cuprates and the FeAs/FeSe layers in the iron-based superconductors (IBS). General similarities in the electronic structures are also found between the \BiS~family and the cuprates/IBS \cite{Yazici2015218}. Electron/hole doping is often necessary to induce superconductivity in IBS, such as oxygen-fluorine (O-F) doping in the well-studied LaFeAsO$_{1-x}$F$_{x}$ \cite{kamihara2008La1111} or hole doping in Ba$_{1-x}$K$_{x}$Fe$_{2}$As$_{2}$ \cite{BaKFe2Asholedoping}. For the \BiS~compounds, electron doping was shown to induce superconductivity through fluorine substitution of oxygen or tetravalent substitution of Lanthanum \cite{YaziciDoping2013PRB}. Also, some members of the \BiS-based superconductors exhibit exotic properties, such as the coexistence of ferromagnetic order and superconductivity in \COFBS~\cite{JieXingPRB.86.214518}. Extensive efforts on studying IBS show that the delicate interplay between magnetism and superconductivity is rather complicated, such as either competition/microscopic coexistence between static antiferromagnetic order and superconductivity in Ba$_{1-x}$K$_{x}$Fe$_{2}$As$_{2}$ \cite{park2009electronic,BaKFe2Asholedoping,paglione2010IBSMag}. Therefore, the \BiS~family presents a new avenue to better understand the underlying physics of lower-dimensional superconductivity, crucial to efforts in uncovering higher $T_{c}$'s.

The first member of the superconducting \BiS-based materials to be discovered was \BOS~\cite{YoshikazuPRB.86.220510}, suggesting that the superconductivity arises from the \BiS~layer. This was confirmed soon after superconductivity was observed in \LOFBS~\cite{YoshikazuJPSJ.81.114725}. Other aspects of the superconductivity in \BiS~family remain unsettled. For example, the superconducting energy gap structure remains unresolved in spite of numerous investigations \cite{Yazici2015218}.

Transverse-field muon spin relaxation (TF-\msr) measurements on high-pressure synthesized (HP) \LOFBS~\cite{YoshikazuJPSJ.81.114725}~find that the temperature dependence of the superfluid density, derived from the measured magnetic penetration depth, $\lambda$, is best described by an anisotropic single-gap $s$-wave model due to two-dimensional Fermi surface nesting at ($\pi,\pi,0$) with strong electronic correlations \cite{PhysRevB.88.180509}. This is also consistent with theoretical work for a single extended $s$-wave band based upon electron-electron correlations~\cite{Liang13}. However, electrical resistivity measurements on ambient-pressure synthesized (AP) \LOFBS~and \COFBS~under applied pressure display behavior consistent with a two-gap model \cite{LaCePressure}. Additionally, TF-\msr~measurements on \BOS~found evidence for multigap superconductivity \cite{Biswas13,Li2013Bi443}. To complicate matters further, theoretical functional renormalization group (FRG) studies on the spin-orbital coupling claim that pairing in the \BiS-based superconductors is a mixture of singlets and triplets \cite{DaiFRG2015}. As the \BiS~family shares similarities with the IBS, and multigap superconductivity has been observed in several IBS, such as Ba$_{1-x}$K$_{x}$Fe$_{2}$As$_{2}$, and Fe$_{1+y}$Te$_{1-x}$Se$_{x}$ \cite{SCinIBSRMP,11122MBSC,Kim2010b,BaKFeAsMBSC}, elucidating the origins of this potential multigap superconductivity in the \BiS~family of superconductors is crucial. We have performed TF-\msr~measurements on \LOFBS~(AP). From the temperature dependence of the superfluid density, we find that \LOFBS~(AP) is well described by a two gap model. However, the anisotropic $s$-wave model cannot be ruled out. Furthermore, an analysis based upon the Uemura relation for unconventional superconductors \cite{UemuraPlot} finds that a number of \BiS-based superconductors conform to this relation, similar to that observed in some IBS and cuprates \cite{LaFeAsOFLuetkensPRL.101.097009,BiswasPRLUemuraPlot13,RenPRLUemuraPlot08}.

\section{Experimental Methods}
TF-\msr~has been widely utilized to probe superconductivity in type-II superconductors at the microscopic level, including the magnetic penetration depth obtained from the muon spin depolarization rate \cite{Sonier00RMP,Sonier07,UemuraPenetrationDepthPRL1992}. 100\% spin-polarized positive muons, each with a momentum of 29.8 MeV/c and kinetic energy of 4.12 MeV, are injected one at a time into the sample in an external magnetic field $H_{ext}$ applied perpendicularly to the initial muon spin polarization. Each muon spin precesses about the local magnetic field $\mathit{B}_{loc}$ at the muon stopping site with the Larmor frequency $\omega$ = $\gamma_{\mu}\mathit{B}_{loc}$, where $\gamma _{\mu}/2\pi$ = 135.53 MHz/T is the muon gyromagnetic ratio. The muons decay with an average life-time of $\tau_{\mu}$ = 2.2 $\mu$s, predominantly emitting a positron along the direction of the muon spin. Measurements of the anisotropic distribution of the decay positrons as well as the lapse time between muon implantation and positron detection for an ensemble of muon decay events yield the time evolution of asymmetry $A(t)$, which is proportional to the muon depolarization.

TF-\msr~experiments on an AP unaligned powder sample of \LOFBS~were carried out in an applied field of 266 Oe in the LAMPF spectrometer at the M20 beamline and in the DR spectrometer at the M15 beamline, TRIUMF, Vancouver, Canada. Details of the synthesis method are described in a previous report~\cite{Yazici13}. Heat capacity measurement gives a single sharp specific heat jump at 2.9 K , with entropy conserved under the superconducting specific heat curve \cite{Yazici13}, evidence of high sample homogeneity. The samples were mounted on a silver holder in the DR spectrometer. The LAMPF spectrometer only requires very thin silver tape to hold the sample. The TF-\msr~data was analyzed with the software \begin{footnotesize}MUSRFIT\end{footnotesize}~\cite{MUSRFIT}.

\section{Results}
\begin{figure}[ht]
 \begin{center}
 \includegraphics[width=0.45\textwidth]{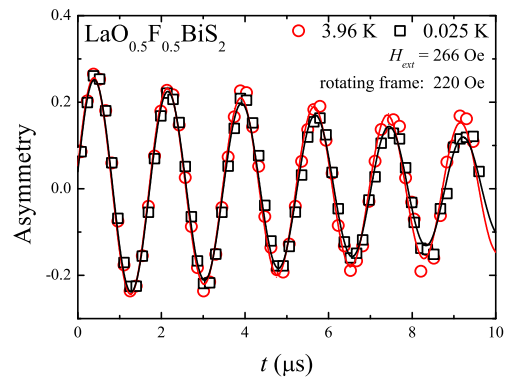}
 \caption{(Color online) TF-\msr~spectra from \LOFBS~(AP) in the normal (circles) and superconducting (squares) states with an external magnetic field of $H_{ext}$ = 266 Oe. Solid curves are fits to the raw data with Eq.~(\ref{eq:AsyLa}). For clarity, the spectra are shown in a rotating reference frame corresponding to a field of 220 Oe~\cite{brewer1994encyclopedia}.}
 \label{fig:Asy}
 \end{center}
\end{figure}

Figure~\ref{fig:Asy} shows the TF-\msr~spectrum in an applied field $H$ of 266 Oe for \LOFBS~at 25~mK (squares) and 3.96~K (circles). For clarity, a rotating reference frame corresponding to a magnetic field of 220 Oe is used to display the TF-\msr~spectrum. Slightly faster damping is observed at base temperature compared to 3.96 K, consistent with an enhanced field inhomogeneity in the vortex state. The Fourier transform of the asymmetry spectrum (not shown here) is not purely Gaussian-shaped; thus, a single Gaussian term along with a background signal does not describe the spectra well. Instead, we find that an additional Lorentzian term along with the Gaussian term is required to best fit the TF-\msr~spectrum, giving the following functional form~\cite{KhasanovVoigtianPRB2008}
\begin{equation}
\begin{split}
\label{eq:AsyLa}
A(t) =& A_{0}\left[f_{s}\exp(-\Lambda t-\textstyle\frac{1}{2}\sigma^2t^2)\cos(\omega_{s} t+\phi)\right.\\*
&+ \left.(1 - f_{s})\mathrm{\mathit{e}}^{-\Lambda_{bg}t}\cos(\omega_{bg} t+\phi)\right],
\end{split}
\end{equation}
where the first and second terms correspond to muons that stop in the sample and silver sample holder, respectively ($f_{s}$ represents the fraction of muons stopping in the sample). The second term is not necessary for the LAMPF spectrometer as no muon stops in the thin sample-holding silver tape. $A_{0}$ is the initial asymmetry of the signal. The Gaussian relaxation rate $\sigma$, which appears below $T_{c}$, is proportional to the rms width of the internal field distribution, which is due to the emergence of flux-line-lattice (FLL) field inhomogeneity in the superconducting state~\cite{Sonier00RMP}. The exponential damping rate $\Lambda$ represents the nuclear dipolar field distribution. The observation of an exponential relaxation rate for a static nuclear dipolar field is unusual. One possible cause of the origin of Lorentzian-like nuclear relaxation $\Lambda$ is the formation of fluorine-$\mu$ or fluorine-$\mu$-fluorine states. However, the typical well-defined shape of precession signals from fluorine-$\mu$-fluorine ``hydrogen~bonding'' \cite{FmuF} is not seen in our TF-\msr~spectrum. The two relaxation terms with the Gaussian rate $\sigma$ and the Lorentzian rate $\Lambda$ are multiplied as the FLL field and nuclear dipole field are completely decoupled. $\omega_{s}$ is the internal precession frequency of muons stopping in the sample, which is used to determine the internal magnetic field. In the DR spectrometer, the background frequency $\omega_{bg}$ and the background relaxation rate $\Lambda_{bg}$ are constant (from the fits $\Lambda_{bg}$ is determined to be $\sim$ 0.0624~(2)~$\mu s^{-1}$). No extra damping component is found in the TF-\msr~spectra down to 25~mK, suggesting no magnetic order in the \LOFBS~(AP).

\begin{figure}[ht]
 \begin{center}
 \includegraphics[width=0.45\textwidth]{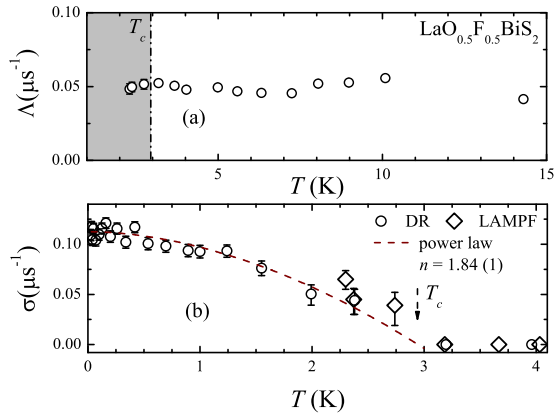}
\caption{(Color online) (a) Temperature dependence of exponential muon relaxation rate $\Lambda$ in \LOFBS~(AP). For all temperatures measured $\Lambda$ displays little temperature dependence. (b) The Gaussian relaxation rate $\sigma$ as a function of temperature. The arrow represents the $T_{c}$ determined from specific heat (Ref. [\onlinecite{Yazici13}]). The dashed dark brown curve is the fit to Eq.~(\ref{eq:SigPow}). Open circles and diamonds represent data taken from the DR and the LAMPF spectrometers, respectively.}
 \label{fig:Sig}
 \end{center}
\end{figure}

The temperature dependences of $\Lambda$~and $\sigma$~obtained from fits of Eq. (\ref{eq:AsyLa}) to the data are given in Figure~\ref{fig:Sig}. $\Lambda$ exhibits a nearly temperature-independent behavior, with an average value of $\Lambda$ = 0.049~(3)~$\mu s^{-1}$ in the normal state. $\Lambda$ is also expected not to change when entering the superconducting state~\cite{LShuPRL.113.166401,SchillingNormalState1982}, and thus is fixed to its normal state average value. A noticeable upturn in $\sigma$ develops below 2.9~K, consistent with $T_{c}$ determined from measurements of heat capacity and electrical resistivity~\cite{Yazici13}.
The temperature dependence of $\sigma$ can then be fit with~\cite{Sonier00RMP}
\begin{equation}
\begin{split}
\label{eq:SigPow}
\sigma (T) = \sigma (0)[1 - (T/T_{c})^{n}], (T < T_{c}),
\end{split}
\end{equation}
yielding $\sigma (0)$ = 0.11~(1)~$\mu s^{-1}$ and $n$ = 1.84~(1) in \LOFBS. The exponent $n < 2$ suggests structure within the superconducting energy gap \cite{Sonier00RMP,Amato97RMP}. Among candidates for this structure are gap nodes \cite{Broholmpowerlawnode,luke1993UPt3node}, multiple gaps \cite{PhysRevLett.88.047002,powerlawmulti}, and $s$-wave anisotropy \cite{takeshita2009aniso}, as we discuss below.

Next we obtain the zero-temperature penetration depth $\lambda(0)$ from the relaxation rate $\sigma (0)$~=~0.11~(1)~$\mu \mathrm{s}^{-1}$. In \LOFBS~(AP) the upper critical field $H_{c2}(0)$ is estimated to be 1.9~T \cite{Awana13}, giving a reduced applied magnetic field $b = H/H_{c2}(0) \approx 0.014 \ll 1$ ($H$ is the applied field). For intermediate values of $b$ (see below), $\sigma(T)$ is related to $\lambda(T)$ by \cite{Brandt03}
\begin{equation}
\begin{split}
\label{eq:Sig}
\sigma(T)/\gamma_{\mu} = A(b)\Phi_0\lambda^{-2}(T),
\end{split}
\end{equation}
where $\Phi_0$ = 2.07 $\times$ $10^{-15}$ Wb is the magnetic-flux quantum, and
\begin{equation}
\begin{split}
\label{eq:Ab}
A(b) = 0.172(1 - b)[1 + 1.21(1 - \sqrt{b})^{3}]/2\pi.
\end{split}
\end{equation}
In our case $A(b)$ = 0.0494 (3), which yields a value $\lambda(0)$ = 0.89 (5) $\mathrm{\mu}$m. Equations~(\ref{eq:Sig}) and~(\ref{eq:Ab}) are valid provided \cite{Brandt03}
\begin{equation}
\begin{split}
\label{eq:kappa}
0.13/\kappa^{2} \ll b \ll 1,
\end{split}
\end{equation}
where $\kappa$ = $\lambda_{ab}(0)/\xi(0)$ is the Ginzburg-Landau parameter \cite{Brandt03}. Here $\lambda_{ab}$ is the in-plane penetration depth and $\xi$ is the Ginzburg-Landau coherence length, given by $\xi(0) = (\Phi_{0}/2\pi H_{c2}(0))^{1/2}$. Since polycrystalline \LOFBS~is a layered compound and is expected to be highly anisotropic, we can estimate $\lambda_{ab}$ from the relation $\lambda = 3^{1/4}\lambda_{ab}$ (Ref.~[\onlinecite{InPlanePen,BarfordPenTheory1988}]). $H_{c2}(0)$ was previously determined to be 1.9 T \cite{Awana13} and using this we obtain $\xi(0)$ = 13.2 nm and $\kappa$ = 46.6. Then  $0.13/\kappa^{2} \approx 6\times10^{-5}$, thus Eq. (\ref{eq:kappa}) is easily satisfied. We note that $A(b)$ is about 20 $\%$ smaller than $A(b=0)$ = 0.0605. The latter estimate is often used, but is only applicable when $b$ is sufficiently small.

The London penetration depth of \BOS~measured by TF-\msr~suggests multigap superconductivity~\cite{Biswas13}, and a two-gap $s$-wave model (the $\alpha$ model) describes the gap structure in \BOS. This model is widely used to characterize many canonical multiband superconductors such as MgB$_{2}$~\cite{PhysRevLett.88.047002}. Applying the same model here, the temperature dependence of $\lambda^{-2}(T)/\lambda^{-2}(0)$ for \LOFBS~is fit by the following functional form within the London approximation~\cite{Tinkham75,KhasanovalphamodelPRL.98.057007,Zurab11}:
\begin{equation}
\begin{split}
\label{eq:LaPen}
\lambda^{-2}(T)/\lambda^{-2}(0) = a\rho[\Delta_{1}(0),T] + (1 - a)\rho[\Delta_{2}(0),T],
\end{split}
\end{equation}
and the equation reverts to the more common form for the isotropic single gap BCS $s$-wave model and the anisotropic $s$-wave model with $a$ = 1. $\rho[\Delta_{i}(0),T]$ is defined by~\cite{Padamsee73}:
\begin{equation}
\begin{split}
\label{LaPen}
\rho[\Delta_{i}(0),T] = 1 + \frac {1}{\pi}\int_{0}^{2\pi}\int_{\Delta_{i}(T,\varphi)}^{\infty}(\frac {\partial f}{\partial E})\frac{EdEd\varphi}{\sqrt{E^{2}-\Delta_{i}(T,\varphi)^{2}}}.
\end{split}
\end{equation}
here $\Delta_{i}(T,\varphi) = \Delta_{i}(0)\delta(T/T_{c})g(\varphi)$, ($\Delta_{i}(0)$ is the energy gap value at $T$ = 0 K for each band, $\delta(T/T_{c})$ is approximated by $1.76\tanh\{1.82[1.018(T_{c}/T-1)]^{0.51}\}$ (Ref.~[\onlinecite{Manzano}])), $g(\varphi)$ is the angular dependence of the gap, and $\varphi$ is the polar angle for the anisotropy. For an isotropic $s$-wave gap, $g(\varphi)$ = 1. For the anisotropic $s$-wave model, $g(\varphi) = (1 + a_{g}\mathrm{cos}(4\varphi)/(1+a_{g})$, with the maximum and minimum gap amplitude ratio $\Delta_{min}/\Delta_{max}$ = $(1 - a_{g})/(1 + a_{g})$ ~\cite{PhysRevB.88.180509}. Finally, $f$ is the Fermi distribution.

\begin{figure}[ht]
 \begin{center}
 \includegraphics[width=0.45\textwidth]{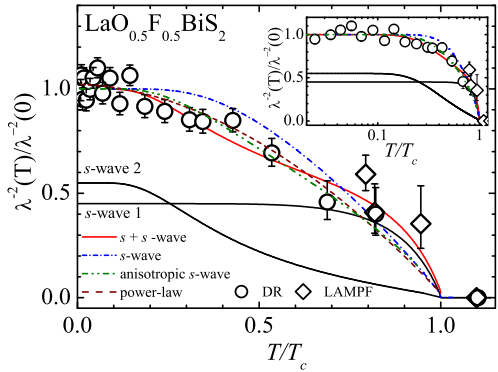}
 \caption{(Color online) Penetration depth plotted as $\lambda^{-2}(T)/\lambda^{-2}(0)$ vs. reduced temperature $T/T_{c}$ for \LOFBS. Curves correspond to two-gap $s + s$ -wave model (solid line), single gap BCS $s$-wave model (dash dot line), anisotropic $s$-wave model (dash dot dot), as well as a power law fit (dashed line). Solid black lines labeled with $s$-wave 1 and $s$-wave 2 represent the individual contributions from the two $s$-wave gaps in the $s + s$ -wave model. The inset shows $\lambda^{-2}(T)$/$\lambda^{-2}(0)$ vs. log$T/T_{c}$ (K), highlighting the quality of fit in the low-temperature regime. Data points taken at the DR and the LAMPF spectrometers are represented by the circle and diamond symbols.}
 \label{fig:WaveFit}
 \end{center}
\end{figure}

The results from the fits of Eq.~(\ref{eq:LaPen}) for a single $s$-wave, anisotropic $s$-wave, and $s + s$~-wave are displayed in Fig.~\ref{fig:WaveFit}. For comparison, a fit using the power-law representing Eq.~(\ref{eq:SigPow}) is also shown (dashed line). The inset shows the data in a semi-log plot to emphasize the quality of fit at low temperatures.
From the fits of $\lambda^{-2}(T)/\lambda^{-2}(0)$, we find that the \LOFBS~(AP) is very well described by both the $s + s$~-wave model and the anisotropic $s$-wave model. In the two-gap model, we determine $\Delta_{1}(0)$~=~0.947~(7)~meV and $\Delta_{2}(0)$~=~0.220~(4)~meV with a weighting factor $a$~=~0.45~(2). From these results, we then determine $2\Delta_{1}(0)/k_{B}T_{c}$~=~7.58~(6) for the large gap, indicating strong coupling, and $2\Delta_{2}(0)/k_{B}T_{c}$~=~1.76~(4) for the second energy gap, below the BCS prediction of 3.74 for weak coupling. This is consistent with theoretical predictions, which find the large band to be strongly coupled and the smaller band to be weakly coupled~\cite{Kresin90}.
The anisotropic $s$-wave model gives $\Delta(0)~=~0.776~(2)$~meV, with $2\Delta(0)/k_{B}T_{c}$~=~4.15~(3) (obtained by averaging gap value over [0,2$\pi$]), which is compatible with the BCS prediction in the weak-coupling limit. The parameters obtained from the different models for representative \BiS-based superconductors are summarized in Table~\ref{tb:FitParameters}.

External pressure applied during synthesis produces a high pressure superconducting phase, with a distinctly higher $T_{c}$~\cite{YoshikazuJPSJ.81.114725,TomitaLaPressure}. Therefore, it is possible that the multigap feature in \LOFBS~(AP) may be due to the coexistence of the ambient pressure phase and smaller amounts of the high pressure phase on the macroscopic level. If this were the case, then evidence of higher $T_{c}$ superconductivity would be observed along with the low-$T_{c}$ phase. Specific heat measurements on the AP sample show no multiple superconductivity features at higher temperature but there is no published data on the HP phase for comparison. There is published magnetic susceptibility data on the HP phase, showing clear evidence of bulk superconductivity at 10 K~\cite{YoshikazuJPSJ.81.114725}, while measurements of the magnetic susceptibility on the AP sample has no obvious transition signal for temperatures higher than $\sim$ 3 K~\cite{Yazici13}. This clearly shows that the AP sample does not contain HP phases, supporting that the possible multigap superconductivity is intrinsic to \LOFBS~(AP).

\begin{table*}[t]
\begin{small}
\begin{center}
\caption{In-plane penetration depth $\lambda_{ab}$ and fit parameters,  $2\Delta(0)/T_{c}$ and $\chi^{2}$ for representative \BiS~superconductors assuming single and two-band $s$-wave energy gaps as well as the anisotropic $s$-wave paring. The parameter $a$ is the weighting ratio of the two $s$-wave gaps. For anisotropic $s$ wave model, gap amplitude ratio is described by $\Delta_{min}/\Delta_{max} = (1 - a_{g})/(1 + a_{g})$. The data for $\mathrm{Bi_{4}O_{4}S_3}$ and \LOFBS~(HP) are from Ref. [\onlinecite{Biswas13}] and Ref. [\onlinecite{PhysRevB.88.180509}], respectively.}
\label{tb:FitParameters}
\centering
\begin{tabular}
{m{2cm}<{\centering}|m{2.3cm}<{\centering}|m{3.8cm}<{\centering}|m{3.8cm}<{\centering}|m{3.8cm}<{\centering} m{0cm}<{\centering}}\hline \hline
 \multicolumn{2}{c|}{} &~\LOFBS~(AP)~&~\BOS~&~\LOFBS~(HP)~& \\ [1ex]\cline{1-6}
 \multicolumn{2}{c|}{$\lambda_{ab}(\rm{\mu m})$} & 0.676 (3) & 0.654 (17) & 0.484 (3) & \\ [1ex]\hline
 \multirow{3}{*}{isotropic $s$} & $\Delta_{0}$ (meV) & 0.374 (5) & 0.88 (2) & 1.47 (3) & \\
 & $2\Delta(0)/k_{B}T_{c}$ & 2.99 (7) & 4.50 (5) & 3.4 (2) & \\
 & $\chi^{2}$ & 2.5 & 1.7 & - & \\ \hline
 \multirow{3}{*}{$s$ $\rm{+}$ $s$ -wave} & $\Delta_{0}$ (meV) & 0.947 (7), 0.220 (4) & 0.93 (3), 0.09 (4) & \multirow{3}{*}{-} & \\
 & $2\Delta(0)/k_{B}T_{c}$ & 7.58 (6), 1.76 (4) & 4.76 (7), 0.44 (9) &  & \\
 & $a$, $\chi^{2}$ & 0.45 (2) , 1.4 & 0.94 (1) , 1.3 & & \\ \hline
 \multirow{3}{*}{anisotropic $s$} & $\Delta_{0}$ (meV) & 0.776 (2) & \multirow{3}{*}{-} & 2.295 & \\
 & $2\Delta(0)/k_{B}T_{c}$$^\dag$ & 4.15 (3) & & 3.74 & \\
 & $a_{g}$ , $\chi^{2}$ & 0.495(1) , 1.55 & & 0.425 , - & \\ \hline\hline
\end{tabular}
\hfill{}
\begin{tablenotes}
  \item[1] $^\dag$Obtained by averaging the gap value over $\varphi$ [0,2$\pi$] (see text).
\end{tablenotes}
\end{center}
\end{small}
\end{table*}

\begin{figure}[ht]
 \begin{center}
 \includegraphics[width = 0.45\textwidth]{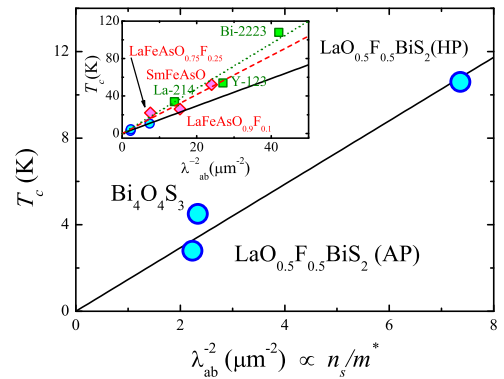}
 \caption{(Color online) The correlations between $T_{c}$ and $\lambda^{-2}_{ab}$ (proportional to the superfluid density $n_{s}/m^{*}$) of the polycrystalline \BiS-based superconductors $\mathrm{Bi_{4}O_{4}S_3}$~\cite{Biswas13}, \LOFBS~(AP), and \LOFBS~(HP)~\cite{PhysRevB.88.180509}. Inset: Uemura plot of the hole-doped iron-based superconductors (diamonds) and high-$T_{c}$ cuprates (squares) as well as the \BiS-based superconductors (circles).}
 \label{fig:UemuraPlot}
 \end{center}
\end{figure}

Figure~\ref{fig:UemuraPlot} shows the linear dependence of $T_{c}$ with $\lambda_{ab}^{-2}$ for representative \BiS-based superconductors which is referred to as the Uemura relation~\cite{UemuraPlot}. $\lambda_{ab}^{-2}$ is proportional to $n_{s}/m^{*}$ (carrier density over effective mass). The slope of the Uemura plot line for \LOFBS~(AP), $\mathrm{Bi_{4}O_{4}S_3}$, and \LOFBS~(HP)~\cite{Biswas13,PhysRevB.88.180509} is 1.47 $\mathrm{K\cdot \mu m^{-2}}$. Similar trends are observed in the iron chalcogenide superconductors such as LaFeAsO$_{1-x}$F$_{x}$, SmFeAsO$_{1-x}$F$_{x}$, and Fe$_{1+y}$Se$_{1-x}$Te$_{x}$, and in many hole-doped cuprate superconductors~\cite{BiswasPRLUemuraPlot13,RenPRLUemuraPlot08,LaFeAsOFLuetkensPRL.101.097009,UemuraPlot}. The \BiS~superconductors conform to the Uemura plot behaving as if they were unconventional. It would be intriguing if more \BiS-based superconductors conform to the Uemura relation.

For many conventional BCS superconductors, the ratio of $T_{c}/T_{F}$ is very small ($T_{F}$ is the Fermi temperature). $T_{F}$ can be obtained from $T_{F} = \varepsilon_{F}/k_{B}$, where Fermi energy $\varepsilon_{F} = n_{s}/m^{*}(\hbar^{2}\pi)$ with $\gamma \varpropto m^{*}$ for two dimensional noninteracting electron gas, and $\varepsilon_{F} \varpropto \sigma^{3/4}\gamma^{-1/4}$ for three dimensional systems (Ref. [\onlinecite{UemuraPlot}]). Here $\gamma$ = 2.53 $\mathrm{mJ~mol^{-1}~K^{-2}}$ is the Sommerfeld coefficient determined by heat capacity measurements~\cite{Yazici13}. This gives a rough estimation of $T_{F}$ of the order of 100 K in \LOFBS~(AP). Interestingly, the ratio $T_{c}/T_{F}$ is larger than for many ordinary BCS superconductors, but close to that for some exotic superconductors including the heavy fermion superconductors $\mathrm{UPt_{3}}$ and $\mathrm{UBe_{13}}$~\cite{UemuraPlot}.

Even though these \BiS~compounds obey the Uemura plot, which is a possible signature of unconventional superconductivity, angle-resolved photoemission spectroscopy measurements on the single crystalline \NOFBS~concluded that it is more likely to be a conventional BCS superconductor mediated by electron-phonon coupling~\cite{NdARPES}. It should be noted that there are many exceptions to the Uemura relation in IBS. For example, a recent study on the iron-based LaFeAsO$_{1-x}$F$_{x}$ system observed the breakdown of the Uemura relation with the application of external pressure~\cite{La1111UemuraBreak}. It is possible that \NOFBS~and other members of the \BiS-based superconductors do not follow the Uemura relation. Future work is necessary to determine if the Nd-member of the \BiS-based family is an exception to the Uemura relation. Additional investigations on energy gap structures in single crystals of the other rare-earth based members would be necessary to better characterize the potential unconventional superconductivity of the \BiS-based layered family.

\section{CONCLUDING REMARKS}
In summary, we have performed TF-\msr~measurements on ambient-pressure synthesized bulk superconducting \LOFBS. From fits to the temperature dependence of the penetration depth, we find \LOFBS~prepared at ambient pressure is well described by the $s + s$~-wave model and the anisotropic $s$-wave model. The $\alpha$ model gives the two supercondcuting gap values of $\Delta_{1}(0)$~=~0.947~meV and $\Delta_{2}(0)$~=~0.22~meV with a weighting factor $a$~=~0.45 for $\Delta_{1}(0)$. The large-gap band is in the strongly coupled limit with $2\Delta_{1}(0)/k_{B}T_{c}$~=~7.58 and the smaller-gap band is weakly coupled with $2\Delta_{2}(0)/k_{B}T_{c}$~=~1.76. Fit using the anisotropic $s$-wave model results in $\Delta(0)$~=~0.776~meV with anisotropic gap amplitude ratio $\Delta_{min}/\Delta_{max}$~=~0.34. Furthermore, \LOFBS~is found to be consistent with the Uemura relation, along with several other \BiS-based superconductors, which is evidence for potential unconventional superconductivity.

\begin{acknowledgments}
We are grateful for the technical assistance from the TRIUMF Centre for Molecular and Materials Science. All the assistance for developing fit codes in {\footnotesize{MATLAB}} are appreciated. This work was supported in part by the National
Natural Science Foundation of China (No. 11474060 and No. 115501107), STCSM of China (No. 15XD1500200), the U.S. National Science Foundation under grant No. DMR-1105380 and DMR-1523588 (CSU-Los Angeles), DMR-1506677 (CSU-Fresno), DMR-0802478 (sample characterization, UCSD), University of California, Riverside, Academic Senate. The research at UCSD was also supported by the U.S. Department of Energy, Office of Basic Energy Sciences, Division of Materials Sciences and Engineering under Grant No. DE-FG02-04ER46105 (sample synthesis).

\end{acknowledgments}
%

\end{document}